\documentstyle[prb,twocolumn,aps]{revtex}

\begin{document}
\tolerance 10000

\draft

\title{Evidence for composite nature of quasiparticles in the 2D
t--J model}

\author{P. B\'eran$^{1,2}$, D. Poilblanc$^{1,3}$ and R. B. Laughlin$^{3}$ }
\vspace*{0.5truecm}

\address{
$^{1}$Laboratoire de Physique Quantique,
Universit\'e Paul Sabatier,
31062 Toulouse, France \\
$^{2}$Institut de Physique, Universit\'e de Neuch\^atel,
CH-2000 Neuch\^atel, Switzerland \\
$^{3}$Department of Physics,
Stanford University, Stanford, California 94305 USA}

\twocolumn[
\date{Received}
\maketitle
\widetext

\vspace*{-1.0truecm}

\begin{abstract}
\begin{center}
\parbox{14cm}{
It is shown that the dynamics of a single hole in a quantum antiferromagnet
(described by the t--J model)
can be simply understood in terms of a composite quasiparticle.
This description provides naturally two different energy scales t and J
corresponding to the inverse masses of the charge (holon) and spin (spinon)
elementary excitations respectively. This picture is consistent with the
exact results obtained on small clusters for the single hole spectral
function and optical conductivity providing that one assumes the existence of
a string-like force of magnitude J between the holon and the spinon.
Then the hole quasiparticle can be interpreted as a bound state of its two
constituents.
}
\end{center}
\end{abstract}

\pacs{
\hspace{1.9cm}
}
]

\narrowtext

\section{Introduction}

The discovery of high-temperature superconductivity has motivated
a considerable effort in the study of strongly correlated fermion systems.
The question of spin-charge separation is an important issue in these systems.
In the case of the 1D t-J model, it is known \cite{Bares} that a hole injected
into the undoped system decays into two elementary excitations,
a charge-1 spinless
excitation (holon) and a neutral spin-1/2 excitation (spinon).
Holons and spinons are independant excitations whose dispersion relations
respectively scale with the hopping integral $t$ and with
the spin-spin coupling constant $J$, so that the decay products
of an injected hole get quickly separated.
The question of spin-charge seperation is more controversed in the case
of the 2D t-J model, which is the simplest model used to describe
the CuO planes of doped $HT_c$ materials.
Anderson \cite{Anderson} suggested that, in analogy with the 1-D case,
a single hole in the 2D t-J model decays into elementary excitations
of spinon- and holon-type.
Other approaches \cite{Schrieffer} describe the single hole
as surrounded by a region with reduced antiferromagnetic order,
leading to a charge-1 spin-1/2 quasiparticle.

Based on exact calculations for small systems
\cite{Dagotto90,Poilblanc93},
we present here evidences that the hole in the 2D t-J model
decays into elementary excitations of spinon and holon type.
Following Ref.\ \onlinecite{Laughlin93}, we make the case that
the Drude conductivity and the total width of the spectral density
of the hole, which are observed \cite{twist,Dagotto92,PZSD}
to depend mainly on parameter $t$, are inconsistent with
the interpretation of the low energy behavior of the spectral density
as a simple quasiparticle whose dispersion scales with $J$.
All these properties of the hole can however
be simply understood in terms of spinons and holons,
{\it provided that, in 2D, there exists a long ranged attraction between
spinons and holons which scales with} $J$.
In the presence of such an attraction, the quasiparticle-like pole seen
in the spectral density \cite{Dagotto90,Poilblanc93} of the single hole
can be regarded	as a bound state of two constituents,
a light one (holon) and a heavy one (spinon).

The long ranged attraction between the heavy constituent
and the light constituent of the hole in 2D can be
visualized using a string picture \cite{Manousakis,Dagotto_mod}.
In this picture, a hole is initially created in an antiferromagnet by removing
an electron.
The hole then hops away, leaving a wake of flipped spins behind.
In absence of spin-flip terms in the Hamiltonian,
the hole is bound to its initial position by the energy cost of the wake
of flipped spins.
A more formal description of this attraction can be obtained
using the gauge theory of quantum antiferromagnets
\cite{Larkin,Zou}, in which the fields of matter describing spinons and holons
are coupled to a {\it confining} gauge field \cite{Laughlin93,Diamantini}.
This leads to the confinement of the constituents of the hole
(spinon and holon), similar to the confinement of quarks which are
the constituents of hadrons and mesons.
The observation of spin-charge separation in the 2D t-J model
at small time and length scales
is therefore important, because it strongly suggests that
elementary excitations with quantum numbers and interaction
similar to those of quarks can be found in quantum antiferromagnets.

In this note, we consider the properties of holes in the t-J model
which is defined by the Hamiltonian
\begin{equation}
{\cal H}={\cal T} + \frac{J}{2} \sum_{<jl>} {\bf S}_j \cdot {\bf S}_l ,
\label{eq:t-J_Ham}
\end{equation}
where the cinetic term ${\cal T}$ is given by
\begin{equation}
{\cal T}= P_G \left[
-t \sum_{<jl> \sigma} c_{j \sigma}^{\dagger} c_{l \sigma}
\right] P_G
\label{eq:cin_Ham}
\end{equation}
and where $P_G$ is the Gutzwiller projector which filters out states
containing doubly occupied sites.
The sum $<jl>$ is performed over near-neighbor pairs, with each pair
counted twice to maintain hermiticity.

\section{Single hole properties}

Let us first consider the propagator of the hole, which is defined by
\begin{equation}
{\cal G}_{k \sigma}(\omega)=
\left< \Psi_0^{(N)} \right|
c_{k \sigma}^{\dagger}
\frac{1}{ \omega + E_0^{(N)} - {\cal H} + i \eta }
c_{k \sigma}
\left| \Psi_0^{(N)} \right> ,
\label{eq:propag_def}
\end{equation}
where $\left| \Psi_0^{(N)} \right>$ denotes the undoped
(N\'eel) groundstate of the system with energy $E_0^{(N)}$,
$
c_{k \sigma} =
N^{-1/2} \sum_j^N e^{ i {\bf k} \cdot {\bf r}_j } c_{j \sigma} .
$
and $N$ is the number of sites.
In Eq. (\ref{eq:propag_def}) and in the following, we set $\hbar=1$.
In Fig.\ \ref{f1}, we show the spectral density
\begin{equation}
A_k(\omega) =
-\frac{1}{\pi} Im {\cal G}_{k \sigma}(\omega)
\label{eq:spectral_density}
\end{equation}
obtained by exact diagonalization of 26-site and 32-site clusters \cite{note1}
for $J=0$ and at different momenta.
At $\Sigma$ ($(\pi/2,\pi/2)$), an $8t$-wide continuum consisting of two lobes
with a hole in the middle is clearly visible \cite{note2}.
When one goes  along the $\Gamma$--M direction (from (0,0) to $(\pi,\pi)$)
weight is re-distributed from negative to positive frequencies.
This is consistent with the calculation of Ref.\ \onlinecite{Laughlin92}
based on chiral spin liquid,
in which these structures result from the decay of the hole into a spinon-holon
pair.
As shown in Fig.\ \ref{f2}(b,c), low energy poles located at the bottom
of the spectrum become visible when the antiferromagnetic coupling $J/t$
is turned on.
They can be interpreted as signatures of quasiparticles \cite{Dagotto90}
with a definite dispersion relation.
Note that, although the quasiparticle-like pole is present for all momenta
(only $\Sigma$ is shown on Fig.\ \ref{f2}) it has a much smaller weight at
momentum $k=(\pi,\pi)$ (M) as we shall discuss later on.

Fig.\ \ref{f2}(d-f) shows the same quantities evaluated using a perturbative
approach \cite{Ruckenstein,KLR} based on linear spin-wave theory.
In this approach, the fermionic operator $c_{k \sigma}$
is expressed as a product of a spin-wave bosonic operator
and a fermionic spinless hole operator.
The hole propagator is then approximated by the propagator
of the spinless hole, which is numerically evaluated
within the self-consistent Born approximation
\cite{Marsiglio,Horsch,Manousakis}.
In this approximation, the line of the spinless hole propagator
is dressed by {\it non-crossing} spin-wave lines.
The perturbative results of Fig. \ref{f2}(d-f) agree well enough with
the exact results of Fig. \ref{f2}(a-c) to help to understand the structure
seen in Fig. \ref{f2}(a-c), as we shall discuss later on.

The quasiparticle dispersion relation vs momentum is shown in
Fig.\ \ref{f3}(a) from an interpolation between the data obtained at
the allowed $\bf K$-values of the 26-cluster.
It agrees remarquably well with the
Green Function Monte Carlo data \cite{Boninsegni} showing in particular
a pronounced minimum at momentum $(\pi/2,\pi/2)$ and a flat band in the
vicinity of $(\pi,0)$. Fig.\ \ref{f3}(b) shows the linear behavior
of the total bandwidth defined as the difference
in quasiparticle energy
between the top of the band at $\Gamma$ (expected to become
exactly degenerate with M in the
thermodynamic limit) and the bottom of the band at $\Sigma$.

In Figs.\ \ref{f4}(a-c), we show the weights
\begin{equation}
Z_{k}=
\frac{ \left|
\left< \Psi_{k}^{(N-1)} \right|
c_{k \sigma} \left| \Psi_0^{(N)} \right>
\right|^2 }{
\left< \Psi_0^{(N)} \right| c_{k \sigma}^{\dagger}
c_{k \sigma} \left| \Psi_0^{(N)} \right>
} ,
\label{eq:Z_def}
\end{equation}
of the quasiparticle poles as a function of $J/t$
for the bottom (Fig.\ \ref{f4}(a)) and the top of the band
(Fig.\ \ref{f4}(b,c)),
where $\left| \Psi_k^{(N-1)} \right>$ denotes
the groundstate of the system with one hole and momentum $k$.
The vanishing of $Z_{k}$ in the limit $J/t \rightarrow 0$
reflects the divergence of the "quasiparticle" size,
i.e. the region of reduced antiferromagnetic order.

At this point, it is important to notice that the spectral density
extends on an energy interval of order $8\, t$ as seen in Fig.\ \ref{f2}.
It is remarkable that this energy width is basically almost independent of
J while, on the other hand, the quasiparticle bandwidth varies linearly
with J. As mentioned previously, the spectral function at
higher energy also shows a strong k-dependence.
These facts are clearly difficult to explain without taking into account
the complex nature of the quasiparticle.

The description of the hole propagator in terms of a simple quasiparticle
dispersing like $J$
is also {\it a priori} \hbox{difficult} to reconcile with the observation
that the
optical conductivity of the hole depends mainly on $t$ as first noticed
by one of us \cite{twist}.
The optical conductivity is defined by
\begin{equation}
\sigma_{xx}(\omega) = 2\pi D \delta(\omega) +
\frac{\pi e^2}{N} \sum_{n \neq 0}
\frac{\left| \left< n \left| j_x \right| 0 \right> \right| ^2}{E_n-E_0}
\delta(\omega-E_n+E_0) ,
\label{eq:opt_cond}
\end{equation}
where the weight $\pi D$ of the Drude peak is simply
proportional to the charge stiffness D given by
\begin{equation}
D = -\frac{e^2 \left< 0 \right| {\cal T} \left| 0 \right> }{4N}
-\frac{e^2}{N} \sum_{n \neq 0}
\frac{\left| \left< n \left| j_x \right| 0 \right> \right| ^2}{E_n-E_0} ,
\label{eq:Drude_weight}
\end{equation}
where $\left| 0 \right>$ and $\left| n \right>$ respectively denote
the groundstate and excited states with energies $E_0$ and $E_n$
of the system with a given number of electrons
and where
\begin{equation}
j_x=it \sum_{l \sigma}
\left[
c_{l \sigma}^{\dagger} c_{l+\hat{x} \, \sigma}
-
c_{l+\hat{x} \, \sigma}^{\dagger} c_{l \sigma}
\right]
\label{eq:para_current}
\end{equation}
with $\hat{x}$ denoting the vector connecting two neighboring sites
along the x direction.

Following Ref.\ \onlinecite{twist} it is convenient to consider the ratio
$D/ e^2 n_h$ since we expect D to scale with the hole doping $n_h$.
This ratio has then the physical meaning of an inverse mass.
Fig.\ \ref{f5}  shows the dependence of the quantity $D/ e^2 n_h$
given in units of $t$
as a function of $J/t$ for a system of 26 sites containing a single hole.
Previous data obtained by averaging over the boundary conditions (to reduce
finite size effects) of
a $4\times 4$ cluster with one or two holes \cite{twist}
are also shown and are in excellent agreement with the
$\sqrt{26}\times\sqrt{26}$ cluster data.
The finite value of $D$ reflects the ability
of the system to conduct electricity \cite{Kohn64}.
Note the small dependence of D in $J$.
It is difficult to see how quasiparticles with properties that depend
on J may lead to a Drude conductivity which depends only on $t$ \cite{twist}.

\section{Spinon-Holon bound state}

We now make the case that the properties of the hole in the 2D t-J model
can be simply explained in terms of elementary excitations
similar to those of the 1D t-J model.
In the case of the 2D t-J model, spinons and holons can be formally defined
within the gauge theory of quantum antiferromagnets \cite{Laughlin93}
as excitations dispersing respectively with $J$ and $t$ and which are
bound together by a {\it string} potential $V({\bf r})=\alpha J |{\bf r}|$,
where $\alpha$ is a dimensionless constant.

The mass of the charged excitation (holon) can be evaluated from
the response to an external electromagnetic field by means
of the {\it f-sum rule} \cite{Kohn64}
\begin{equation}
\int_0^{\infty} d\omega \sigma_{xx}(\omega) =
\pi \frac{n e^2}{2m} ,
\label{eq:f-sum_rule}
\end{equation}
which relates the total weight of optical conductivity
to the effective mass $m$ and to the density $n$ of charge carriers.
Using Eq. (\ref{eq:f-sum_rule}), the quantity $1/2m$ can be expressed
as a sum of two terms, (i) $D/ e^2 n_h$ shown in Fig.\ \ref{f5}
(called ``inverse optical mass'' in Ref.\ \onlinecite{twist})
and (ii) the {\it finite frequency} integrated weight
$\int_{0^+}^{\infty}\sigma_{xx}(\omega) \, d\omega / (\pi e^2 n_h)$
also shown in Fig.\ \ref{f5}.
This leads to
\begin{equation}
m \cong 0.7 \frac{1}{t}
\label{eq:holon_mass}
\end{equation}
in units where the lattice bound length is set equal to one,
i.e. the holon mass depends mainly on parameter $t$.
Note that this estimate for the holon mass agrees well
with that of chiral spin liquid theory \cite{Laughlin93}.
Both Eqs. (\ref{eq:f-sum_rule}) and (\ref{eq:holon_mass})
are consistent with exact results \cite{twist,Dagotto92} for systems
with several holes, in which the total weight is
found to be nearly independent of $J$ and proportional to doping.

The quasiparticle-like pole appearing in the spectral density
of Fig.\ \ref{f2} for finite $J$ can be attributed
to the spinon-holon bound state.
The dispersion of the spinon-holon pair is characterized
by the dispersion of the heaviest object,
which is the spinon in the case of interest $J \approx 0.2 t$.
This is consistent with the finding that the quasiparticle
bandwidth scales with $J$, shown in Fig.\ \ref{f3}.
Also, the vanishing of the quasiparticle weight $Z$
in the limit $J \rightarrow 0$ seen in Fig.\ \ref{f4}
reflects the divergence of the size of the spinon-holon bound state,
consistent with the vanishing of the confining potential
$V({\bf r})=\alpha J |{\bf r}|$ in this limit.

As can be seen in Fig.\ \ref{f2}(e,f), the spectral density evaluated
perturbatively for finite $J$ shows a series of resonances
above the quasiparticle-like pole.
These so-called string resonances \cite{Manousakis} can be interpreted
as higher energy bound states of the spinon-holon pair.
The energy dependence of both the quasiparticle-like pole
and the first higher-energy resonance is displayed in
Fig.\ \ref{f6}.
In this figure, the straight lines correspond to the fit
\begin{equation}
E_n = \beta t + \gamma_n (J/t)^{2/3} t ,
\label{eq:string_power_low}
\end{equation}
where $\beta=-3.28$, $\gamma_0=2.16$ for the quasiparticle-like pole
and $\gamma_1=5.46$ for the first higher-energy resonance.
This is the behavior expected \cite{Manousakis,Laughlin93}
for the spectrum of a light particle in orbit around a heavy one,
described by the Hamiltonian
\begin{equation}
H = -\frac{1}{2m} \nabla^2 + \alpha J |{\bf r}| + \beta t ,
\label{eq:string_Hamiltonian}
\end{equation}
where $m$ is the mass of the light particle (holon)
estimated using the f-sum rule.
Using the energy dependence of either the quasiparticle-like pole
or the first higher-energy resonance,
one respectively finds $\alpha=1.64$ or $\alpha=2.16$
for the string tension.
Exact results \cite{Poilblanc92} for the quasiparticle-like pole
are also well fitted using Eq.(\ref{eq:string_power_low})
with constants $\beta=-3.359$ and $\gamma_0=2.77$,
leading to $\alpha=2.39$.
Nevertheless the observation of high-energy resonances in exact
diagonalizations of small clusters has been more controversial.
Although some authors \cite{Dagotto90} have attributed some peaks in the
spectral function of the $4\times 4$ cluster to
these resonances, similar studies carried out on
larger clusters \cite{Poilblanc93} for $J\ge 0.3$  failed to detect any sharp
structure.
However, it is tempting to attribute the structure appearing
at $\omega \approx -2.3 t$ in Fig.\ \ref{f2}(b)
and at $\omega \approx -1.7 t$ in Fig.\ \ref{f2}(c)
(indicated by arrows on the plot) to the first higher-energy string resonance.
This structure can be fitted using Eq.(\ref{eq:string_power_low})
with constants $\beta=-3.359$ and $\gamma_1=4.95$,
leading to $\alpha=1.86$.
Note that these data correspond to quite small J/t ratios.
For increasing J/t (let's say $J/t>0.2$) the lifetimes of the string states
becomes rapidly too small to be observable.

As seen in Figs.\ \ref{f1} and \ref{f2} significant spectral weight appears
on a broad $8t$-wide energy range. This continuum can be crudely viewed
as the excitation spectrum of the light particle (holon) whose oscillator
strength lies mainly at high frequencies.

We conclude by a comment on the rather weak dependence of the Drude weight
on $J$
shown in Fig.\ \ref{f5} for one or several holes in small t-J clusters.
This implies that a hole doped into the system has a dynamic mass
which depends on $t$ only.
This is in contradiction with the linear behavior of the
quasiparticle bandwidth as a function of $J$ seen in Fig.\ \ref{f3}(b),
which implies that the mass of the hole depends on $J$ only.
It is very likely that one of these results is contaminated by finite
size effects but, so far, it is impossible to determine
which one without computations on larger samples.
It is however important to settle the question of which of these
results holds in the thermodynamic limit
in order to determine wether or not holes decay
into more elementary constituents at low doping:
If our observation that the Drude weight is independent of $J$ is
not seen on larger systems,
this would mean that this decay is indeed not possible.
This would be the case if the Drude ``pole'' separates, in the
thermodynamic limit, into a legitimate Drude pole containing a small
fraction $\propto J$ and a very low frequency band containing the rest.

\section{Acknowledgments}

DP wishes to thank the Stanford University for its hospitality during the early
stage
of this work and also acknowledges support from the EEC Human Capital and
Mobility
program under Grant No. CHRX-CT93-0332. DP and PB acknowledge IDRIS
(Orsay, France) for allocation of CPU time on the C94 and C98 CRAY
supercomputers.

%
%
\begin{figure}
\caption{
Single hole spectral functions for J=0 calculated on small 2D clusters
with 32 and 26 lattice sites. The $\Gamma$, M and $\Sigma$ points in
reciprocal space corresponds to momenta (0,0), $(\pi,\pi)$ and
$(\pi/2,\pi/2)$ respectively as shown in the insert.
The spectral function at $(\pi/2,\pi/2)$
is in fact obtained by averaging the two spectral functions taken at the
two nearest momenta
\protect\cite{note2}.
}
\label{f1}
\end{figure}

%
%
\begin{figure}
\caption{
Spectral function at $\Sigma$
\protect\cite{note2}
for various J/t values calculated on the same
26-site cluster. (a-c) corresponds to the exact results and (d-f) to the
spin-wave calculations.
In (b,c) arrows indicate the location of the first string resonance.
The dashed curves in panel (d-f) correspond
to a $40 \times 40$-site cluster, for which finite size effects are
negligible.
}
\label{f2}
\end{figure}

%
%
\begin{figure}
\caption{(a) Quasiparticle dispersion along some symmetry directions of the
Brillouin zone obtained by interpolating exact results for a 26-sites cluster (
full line )
and obtained from spin-wave calculation for a $40 \times 40$-sites cluster
(dashed line).
(b) Quasiparticle bandwidth vs J/t calculated on a  26-site cluster.
Hexagons and triangles respectively denote exact and perturbative results.
A linear behavior like 2.2 J/t is also shown for comparison.}
\label{f3}
\end{figure}

%
%
\begin{figure}
\caption{
Quasiparticle weights vs J/t calculated on a 26-site cluster for
various location in k-space.
(a), (b) and (c) correspond to the $\Sigma$,
$\Gamma$ and M points as indicated on the figures.
Filled and empty squares respectively correspond to exact and perturbative
results.}
\label{f4}
\end{figure}

%
%
\begin{figure}
\caption{
$D/e^2 n_h$ for a single hole on a 26-site cluster (filled hexagons)
in units of $t$.
The finite frequency integrated weight
$\int_{0^+}^{\infty}\sigma_{xx}(\omega) \, d\omega / (\pi e^2 n_h)$
is also shown (empty hexagons).
Similar results are given for one (triangles) or two (squares) holes
on a $4\times 4$ cluster \protect\cite{twist}.
In that case an average over the boundary conditions was used,
leading to values larger than those obtained in Ref. \protect\cite{Dagotto92}.
}
\label{f5}
\end{figure}

%
%
\begin{figure}
\caption{
Ground-state energy for a single hole as a function of $J$,
obtained from exact calculations for a 26-site cluster (filled triangles)
and from spin-wave calculations (empty triangles).
Similar results are given for the energy of the first string resonnance
(squares) at momentum corresponding to the band bottom
of the quasiparticle-like pole.
In the case of exact results, the first string resonnance
is identified as the structure in the spectral density visible
at $\omega=-2.3t$ in Fig.\ \protect\ref{f2}(b) and at $\omega=-1.7t$
in Fig.\ \protect\ref{f2}(c).
}
\label{f6}
\end{figure}


\begin{references}
\bibitem[16]{note2}
On finite tilted clusters $\Sigma$ is in fact slighly shifted
from $(\pi/2,\pi/2)$. This is a finite size effect due to
the constraint of tilted periodic boundary conditions
\protect\cite{Poilblanc93}.
\end{references}
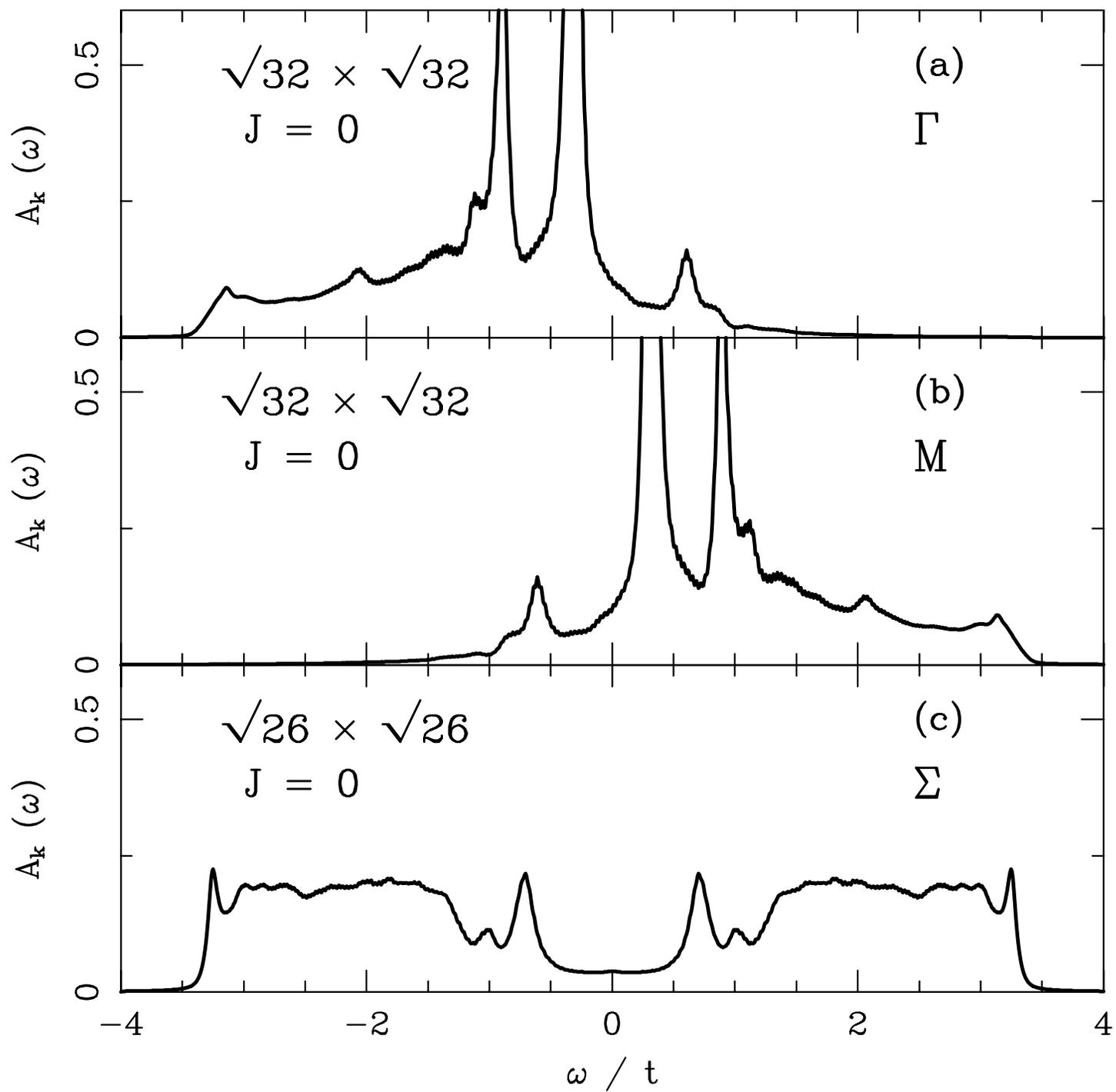
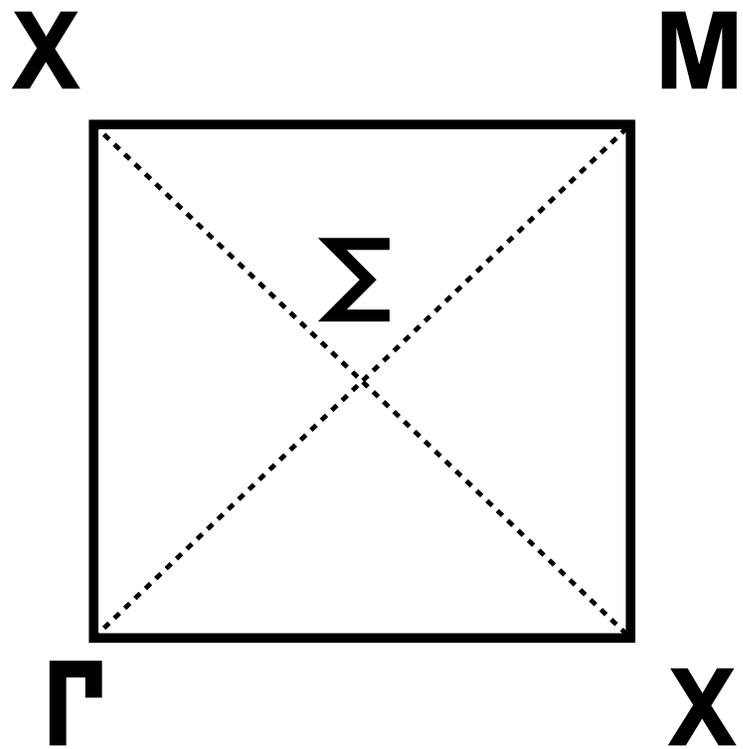
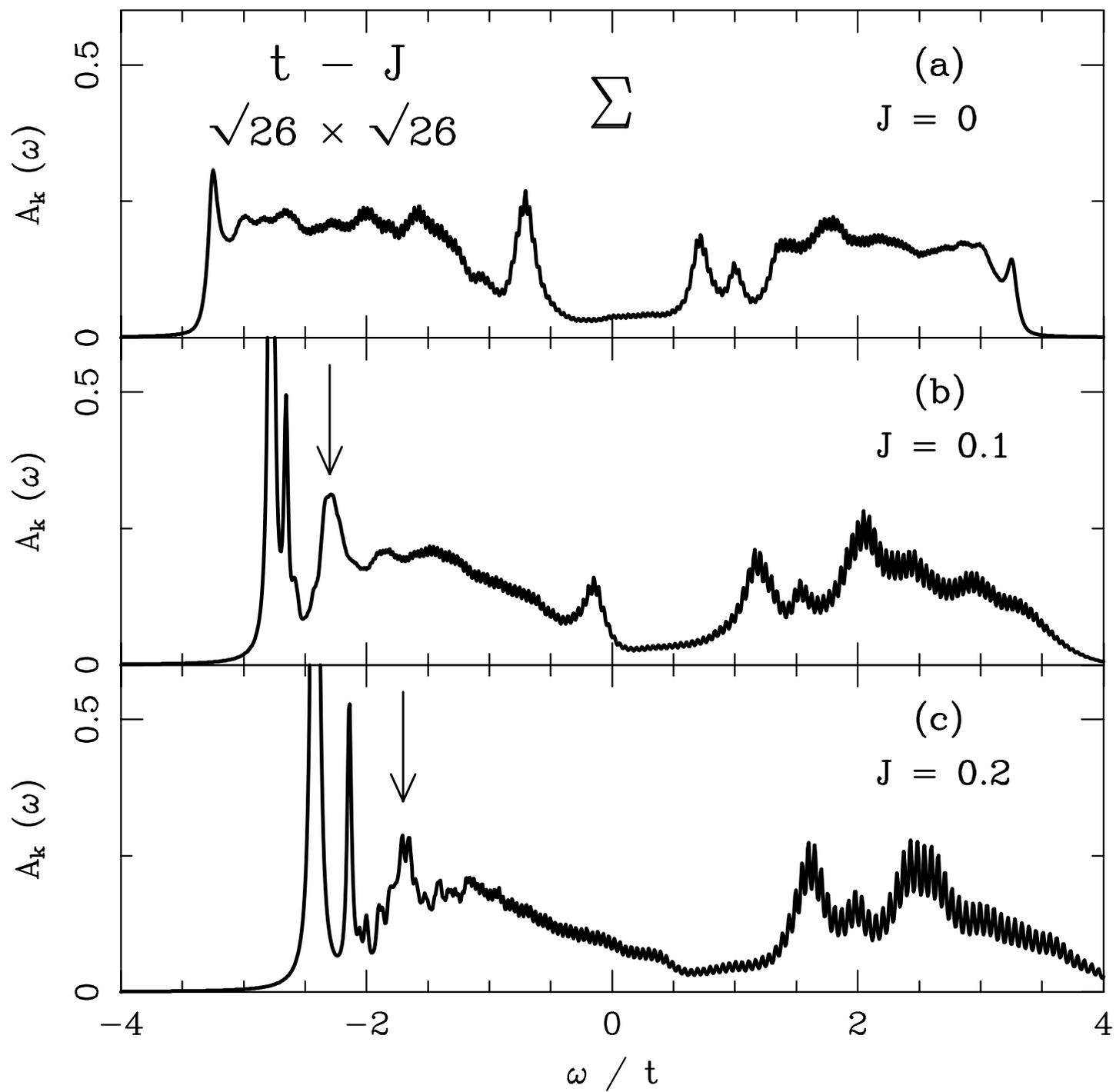
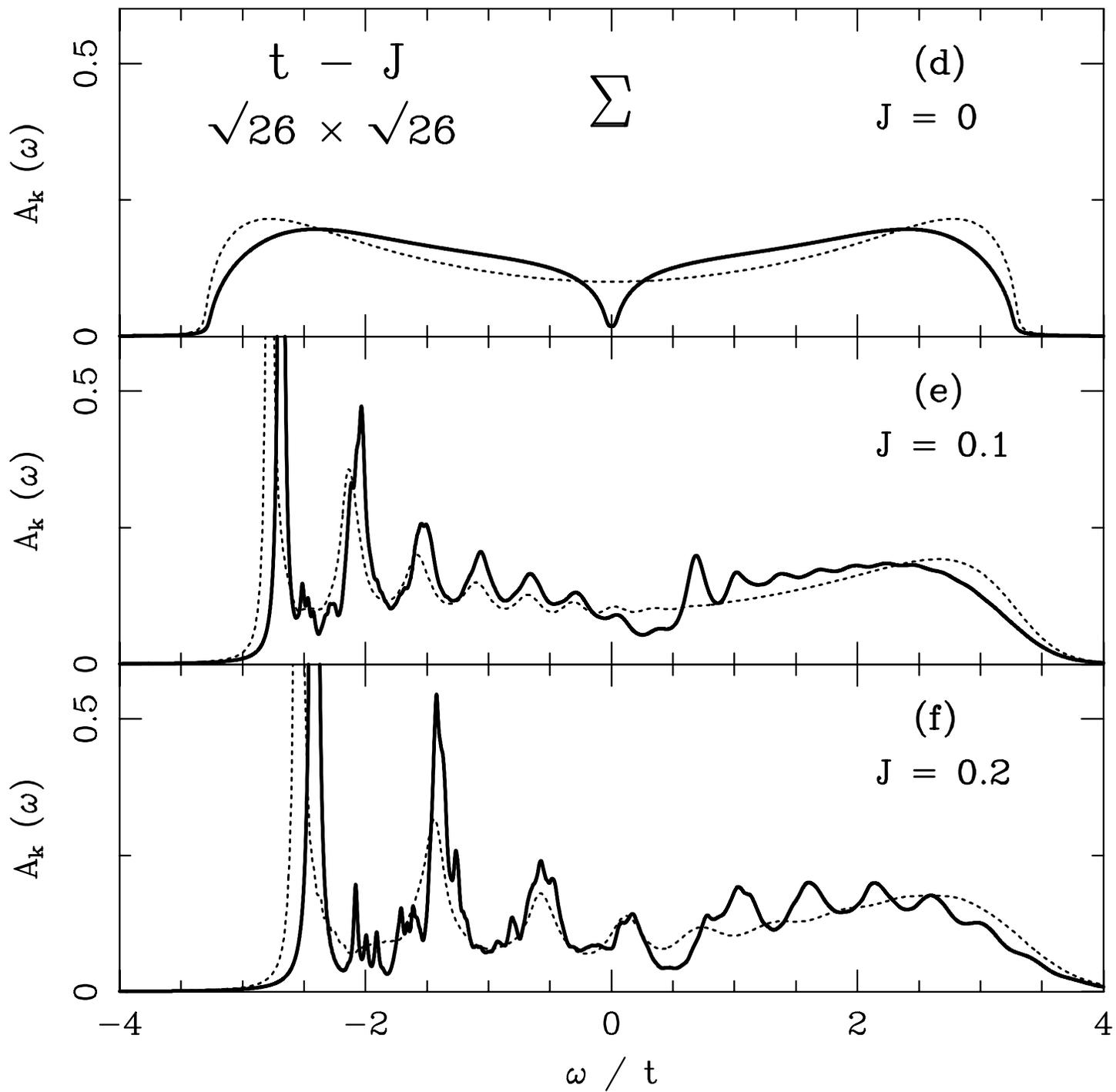
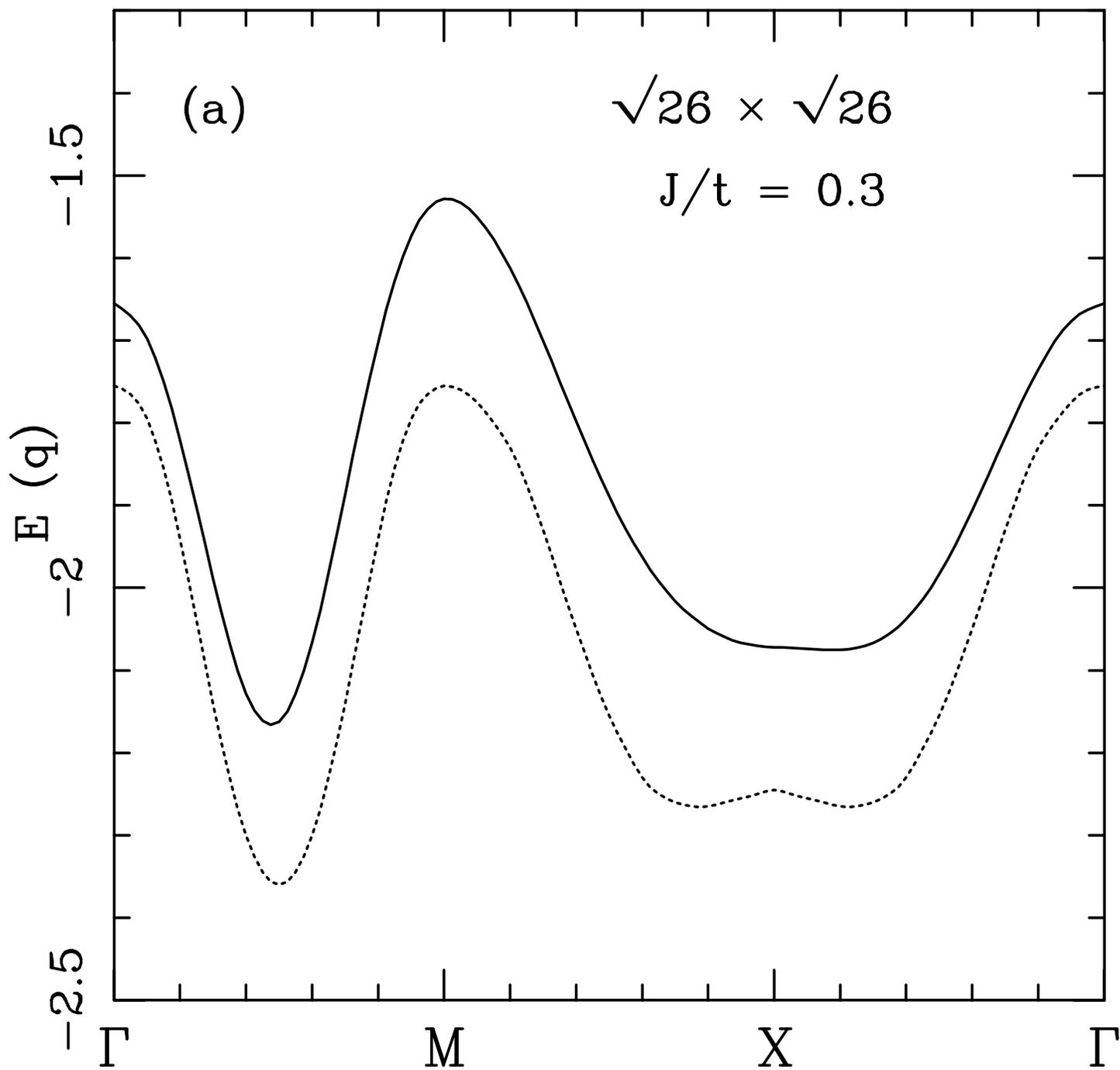
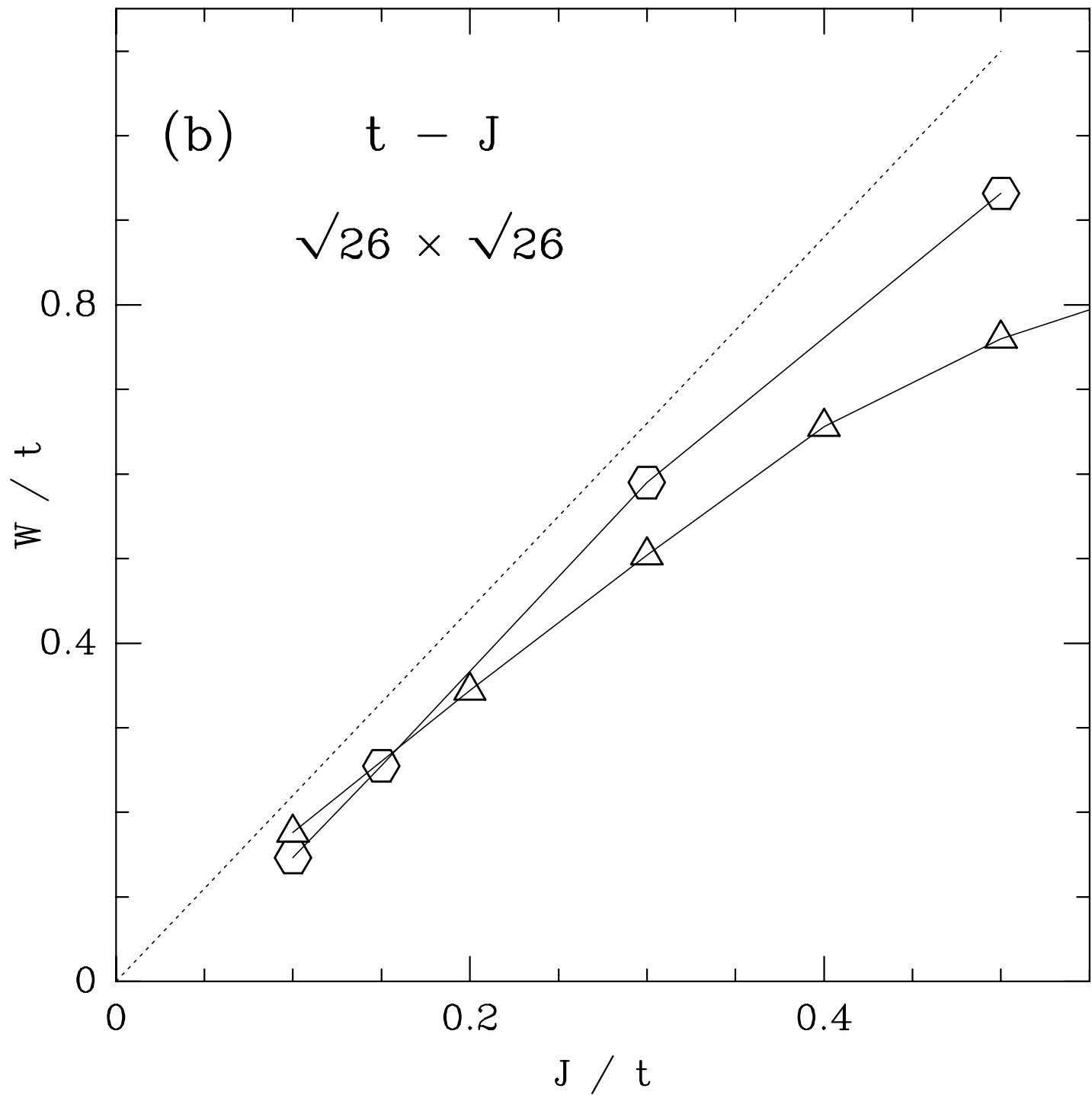
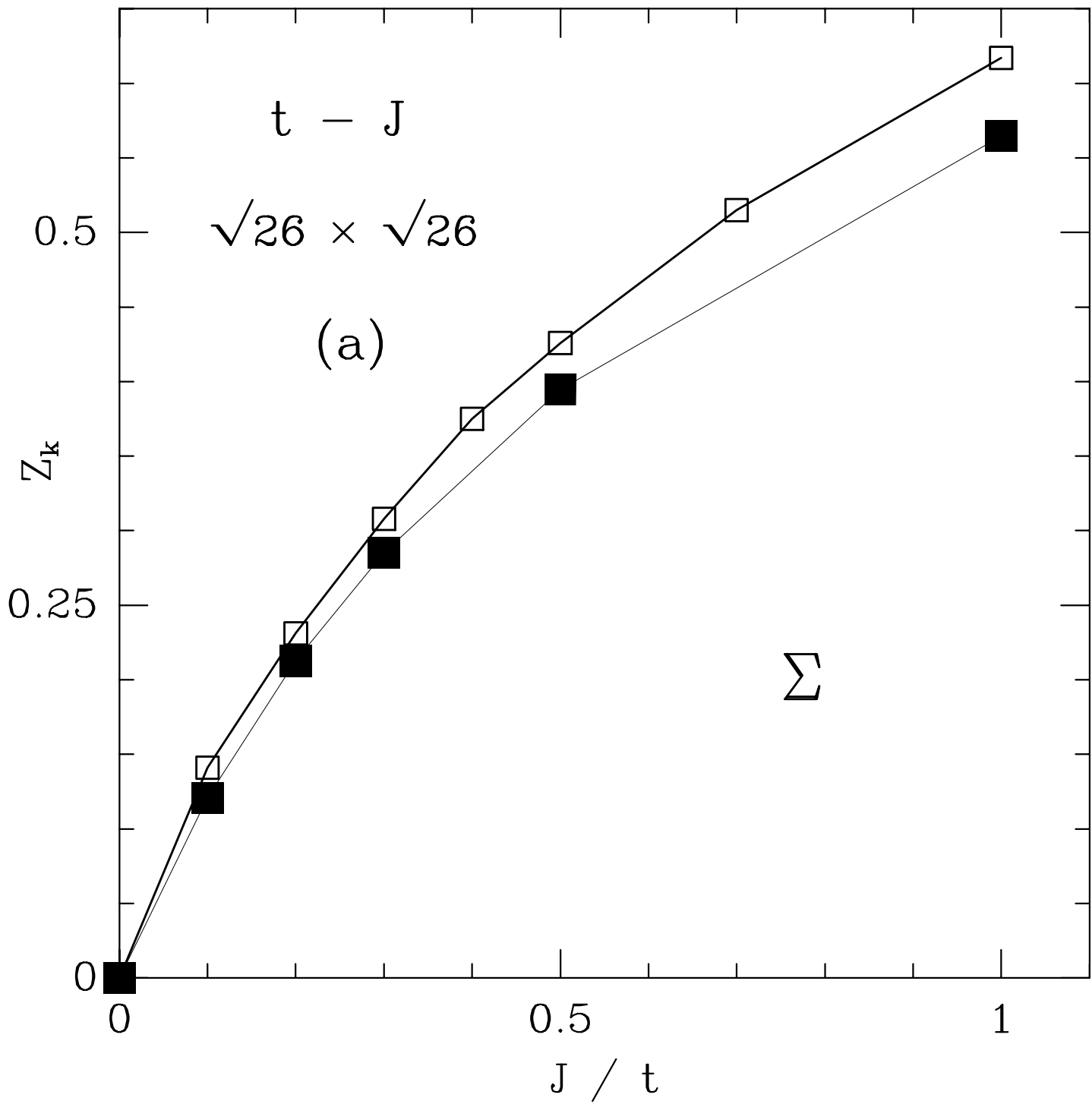
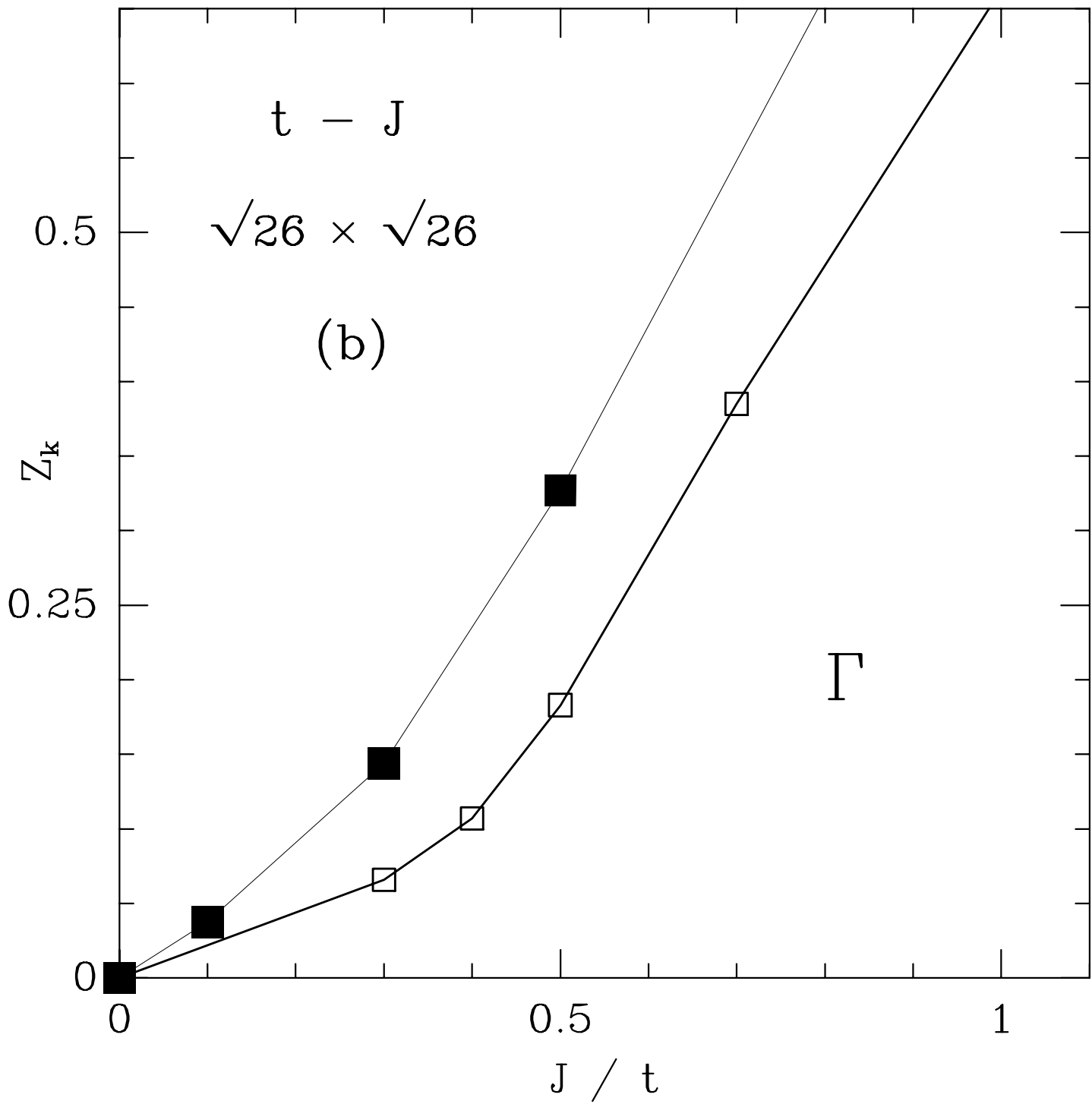
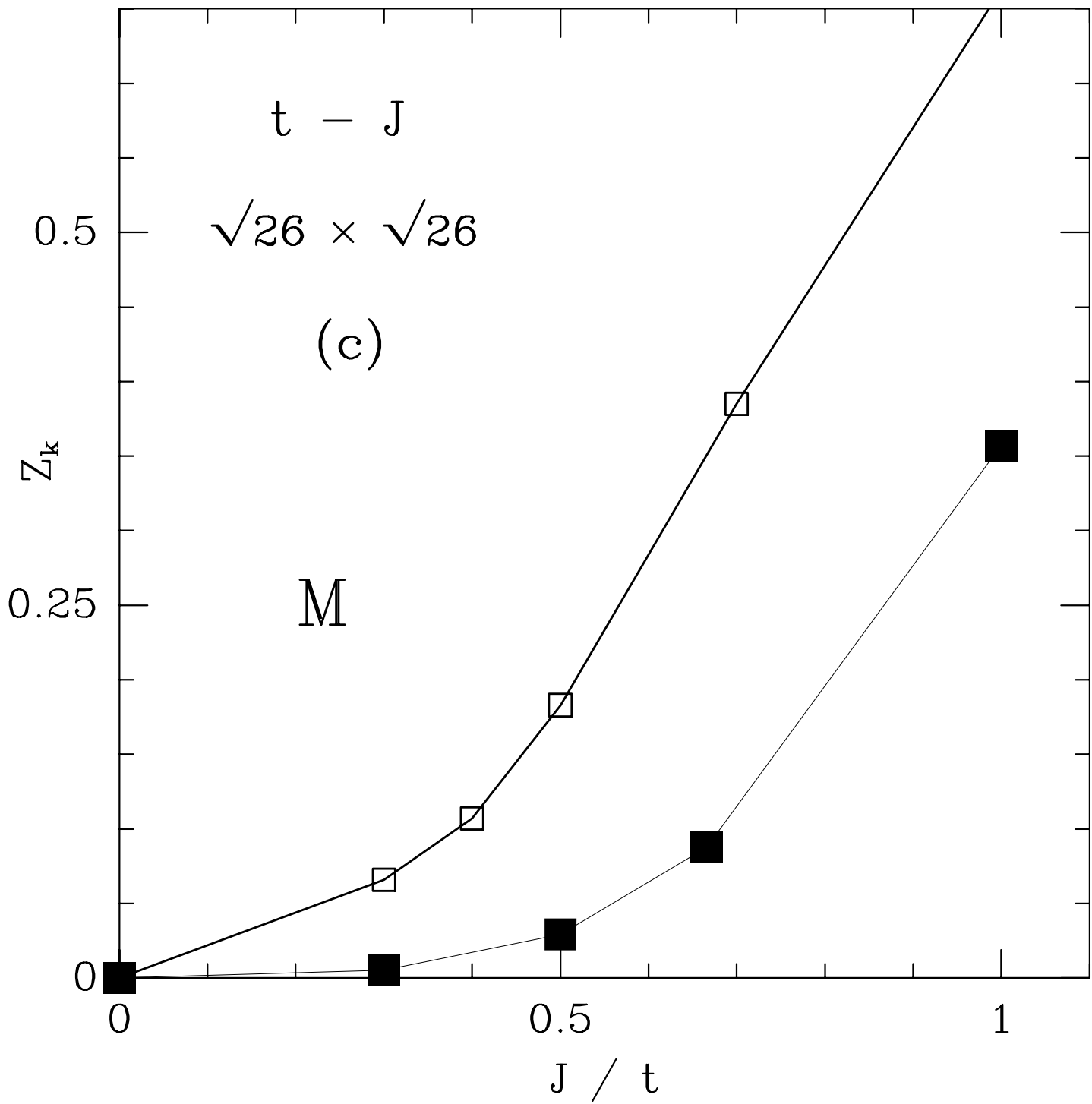
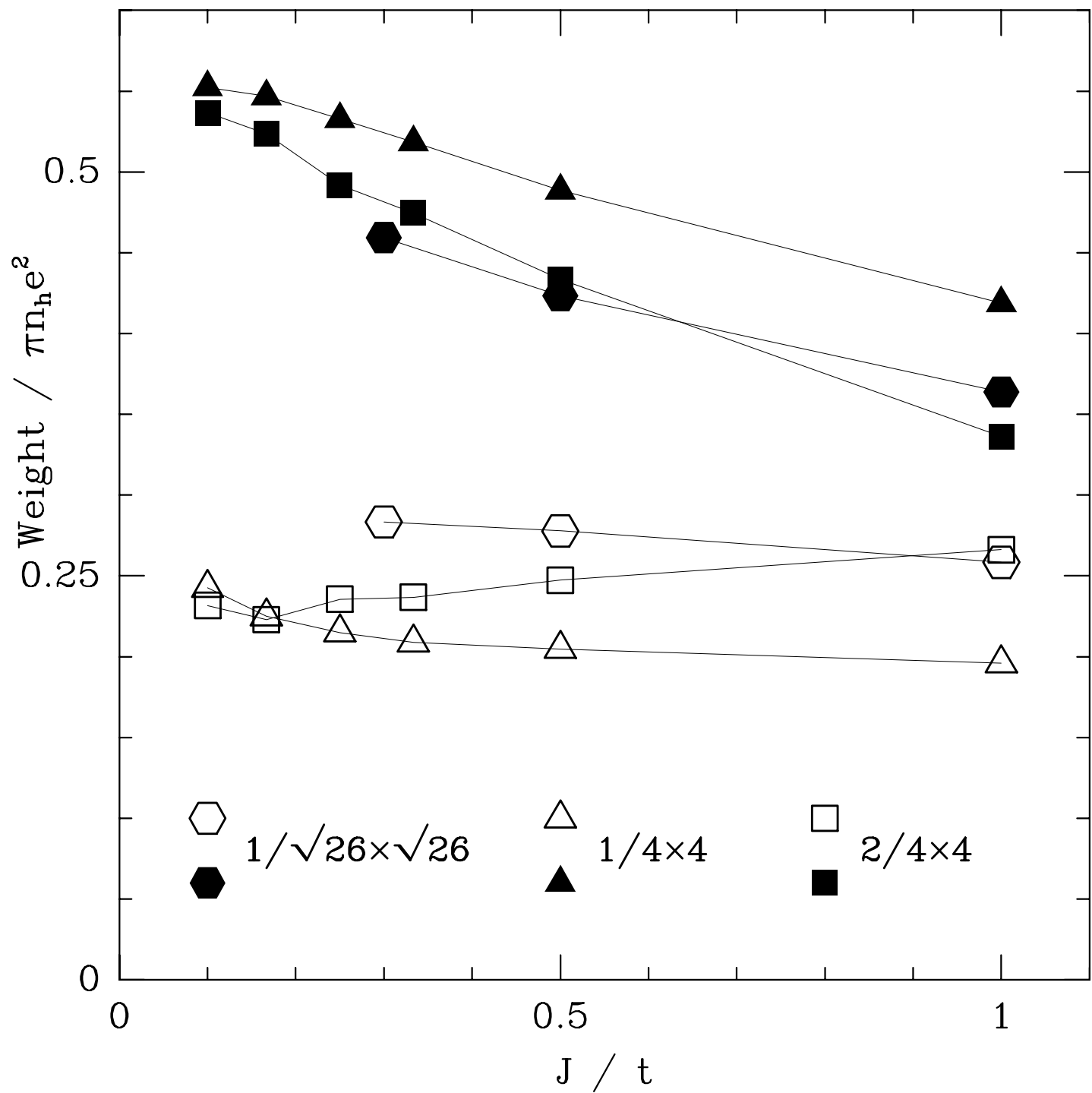
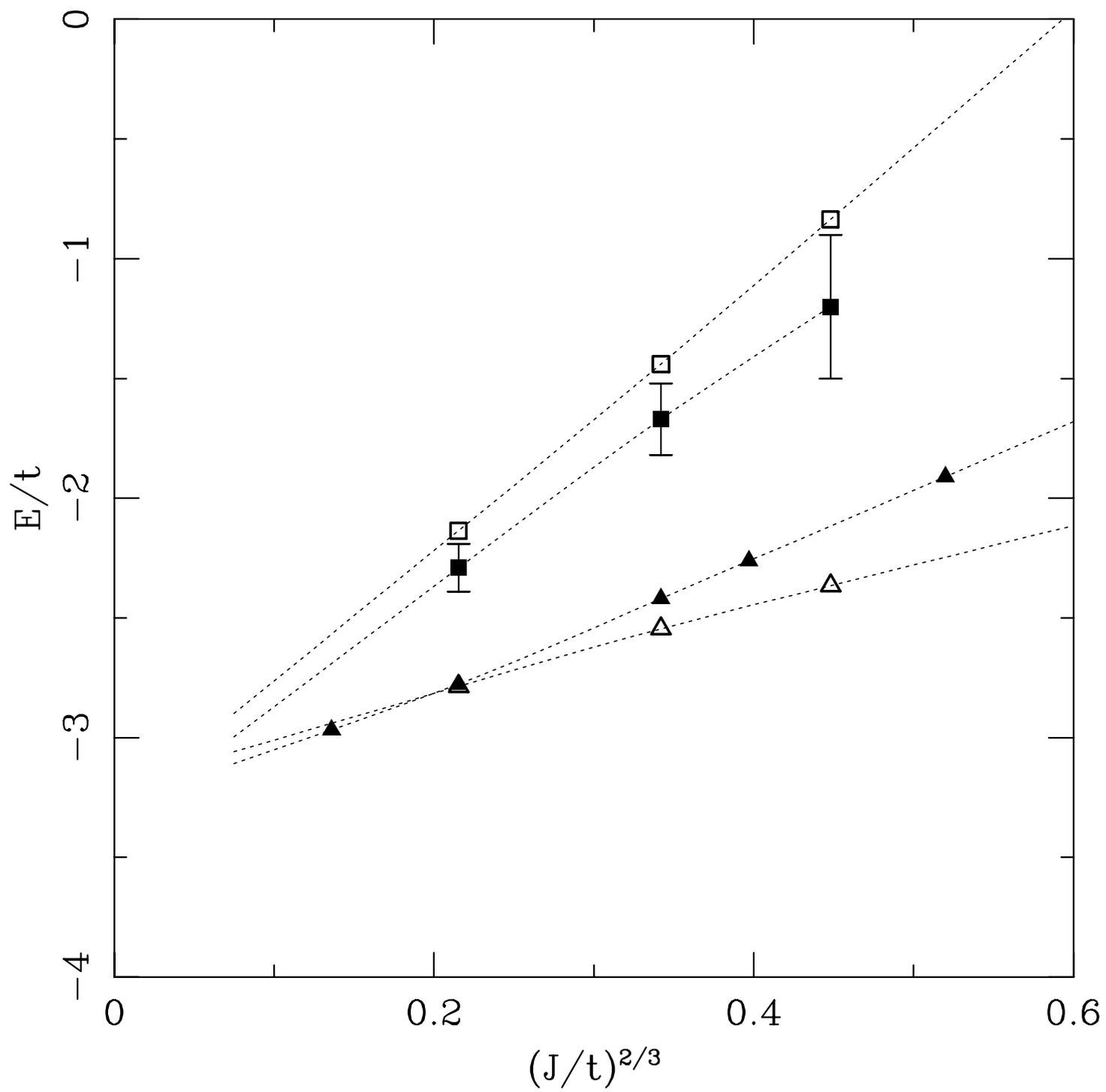
\end{document}